\documentclass[twocolumn,doublespacing,showpacs,preprintnumbers,amsmath,amssymb]{revtex4-1}
\usepackage{amssymb}
\usepackage{txfonts}
\usepackage{graphicx}
\usepackage{subfigure}
\usepackage{dcolumn}
\usepackage{bm}
\usepackage{sidecap}

\begin{document}

\title{Continuous-variable measurement-device-independent quantum key distribution \\ using squeezed states}
\author{Yi-Chen Zhang$^1$}
\author{Zhengyu Li$^{2,3}$}
\author{Song Yu$^{1,}$}
\email{yusong@bupt.edu.cn.}
\author{Wanyi Gu$^1$}
\author{Xiang Peng$^2$}
\author{Hong Guo$^{1,2,3}$}

\affiliation{$^1$State Key Laboratory of Information Photonics and Optical Communications, Beijing University of Posts and Telecommunications, Beijing 100876, China }

\affiliation{$^2$State Key Laboratory of Advanced Optical Communication Systems and Networks, School of Electronics Engineering and Computer Science, Center for Quantum Information Technology, Peking University, Beijing 100871, China}

\affiliation{$^3$Center for Computational Science and Engineering, Peking University, Beijing 100871, China}

\date{\today}

\begin{abstract}
A continuous-variable measurement-device-independent quantum key distribution (CV-MDI QKD) protocol using squeezed states is proposed where the two legitimate partners send Gaussian-modulated squeezed states to an untrusted third party to realize the measurement. Security analysis shows that the protocol can not only defend all detector side channels, but also attain higher secret key rates than the coherent-state-based protocol. We also present a method to improve the squeezed-state CV-MDI QKD protocol by adding proper Gaussian noise to the reconciliation side. It is found that there is an optimal added noise to optimize the performance of the protocol in terms of both key rates and maximal transmission distances. The resulting protocol shows the potential of long-distance secure communication using the CV-MDI QKD protocol.
\end{abstract}

\pacs{03.67.Dd, 03.67.Hk}
\maketitle


\section{Introduction}

Quantum key distribution (QKD)~\cite{Gisin_RevModPhys_2002,Scarani_RevModPhys_2009} is the most prominent application of quantum information science, which accomplishes the secure key distribution phase of an encrypted communication between two legitimate partners, i.e., Alice and Bob. The continuous-variable approach of QKD (CV-QKD)~\cite{Braunstein_RevModPhys_2005,Xiang-Bin_PhysReport_2007,Weedbrook_RevModPhys_2012}, based on the Gaussian modulation of Gaussian states, has attracted much attention in the past few years~\cite{Weedbrook_RevModPhys_2012,Madsen_nature_2012,Jouguet_nature_2013,Weedbrook_PhysRevA_2013} mainly because its associated homodyne or heterodyne detection offers the prospect of high detection efficiency and high repetition rate. Generally speaking, there are eight types of one-way Gaussian CV-QKD protocols, which are classified according to Alice's sending states (squeezed or coherent states), Bob's measurement methods (homodyne or heterodyne measurement), and reconciliation methods (direct or reverse reconciliation )~\cite{Ralph_PhysRevA_1999,Hillery_PhysRevA_2000,Cerf_PhysRevA_2001,Gottesman_PhysRevA_2001,Grosshans_PhysRevLett_2002,Grosshans_Nature_2003,Weedbrook_PhysRevLett_2004,Patron_PhysRevLett_2009}.

A CV-QKD protocol using squeezed states, heterodyne detection, and reverse reconciliation~\cite{Patron_PhysRevLett_2009} outperforms these eight Gaussian protocols~\cite{Patron_PhysRevLett_2009,G_Patron_PhD_2007}, which can be treated as the protocol using squeezed states and homodyne detection~\cite{Ralph_PhysRevA_1999,Hillery_PhysRevA_2000,Cerf_PhysRevA_2001,Gottesman_PhysRevA_2001} followed by a Gaussian added noise. Furthermore, a trusted added noise model is introduced into the squeezed state and homodyne detection protocol, and it performs a longer transmission distance with carefully chosen noise parameters~\cite{Pirandola_PhysRevLett_2009,Patron_PhysRevLett_2009,G_Patron_PhD_2007}. The increased performance can be understood as that adding trusted noise to the receiver will cause the mutual information between the eavesdropper and the receiver to decrease more than that between the two legitimate partners.

Comparing to the generation of a coherent state, generating a squeezed state is much more difficult, which becomes the most difficult part of implementing the CV-QKD protocol using squeezed states. Recent research shows that the largest achievable two-mode squeezing in a stable optical configuration has already been reached at about $10$~dB~\cite{Eberle_OptExpr_2013}, and an experiment using the CV-QKD protocol with squeezed states and homodyne detection has been successfully demonstrated~\cite{Madsen_nature_2012,Peuntinger_arXiv_2014}, which shows the potential for implementing the squeezed-state CV-QKD in real life.

In this paper, we introduce the use of squeezed states into a recently proposed protocol, the continuous-variable measurement-device-independent QKD (CV-MDI QKD) protocol~\cite{Zhengyu_PhysRevA_2013,Pirandola_arXiv_2013}. The MDI-QKD protocol~\cite{Pirandola_PhysRevLett_2012_MDI,Lo_PhysRevLett_2012_MDI,XFMa_PhysRevA2_2012} was first proposed to defend detector side channels. Then many methods were introduced to improve the secret key rate and transmission distance of the protocol, such as using different kinds of sources~\cite{XBWang_PhysRevA_2013,Zhou_PhysRevA_2013,Li_OpticsLetters_2014}, explicitly utilizing the decoy states to dramatically increase the secret key rates~\cite{XBWang_PhysRevA2_2013,XBWang_PhysRevA_2014}, enhancing the practical security by tight finite-size analysis~\cite{XFMa_PhysRevA_2012,TTSOng_PhysRevA_2012,Curty_NatCommun_2014}, etc. All the efforts were aimed at improving the performance of the protocol, which are also our pursuits on the CV-MDI QKD protocol. Here we first present the equivalent entanglement-based (EB) scheme and the prepare-and-measure (PM) scheme of the squeezed-state CV-MDI QKD. We find that the transmission distance of the squeezed-state CV-MDI QKD protocol is longer than the coherent-state-based protocol. In addition, we introduce the trusted added noise model to the receiver, which can further improve the transmission distance. Furthermore, in the most asymmetric case, even if the variance in the EPR is as small as $5.04$ (referring to $10$~dB squeezing~\cite{Text}), the transmission distance can still reach $100.5$~km, which shows the potential for actually implementing the CV-MDI QKD protocol using squeezed states.

The rest of this paper is organized as follows. In Sec.~\ref{sec:2}, we propose a CV-MDI QKD protocol using squeezed states. We optimize the CV-MDI QKD protocol using squeezed states by adding optimal Gaussian noise to the reconciliation side in Sec.~\ref{sec:3}. In Sec.~\ref{sec:4}, we show the numerical simulation results of the secret key rate and give the optimal value of the added noise under different situations. Our conclusions are drawn in Sec.~\ref{sec:5}.

\section{\label{sec:2} SQUEEZED-STATE CV-MDI QKD }

In this section, we first present the idea and basic notions of the CV-MDI QKD protocol using squeezed states and then derive the secure bound of the protocol. The standard PM description of the CV-MDI QKD protocol using squeezed states is shown in Fig.~1 and is described as follows:

\begin{figure}[t]
\centerline{\includegraphics[width=8.0cm]{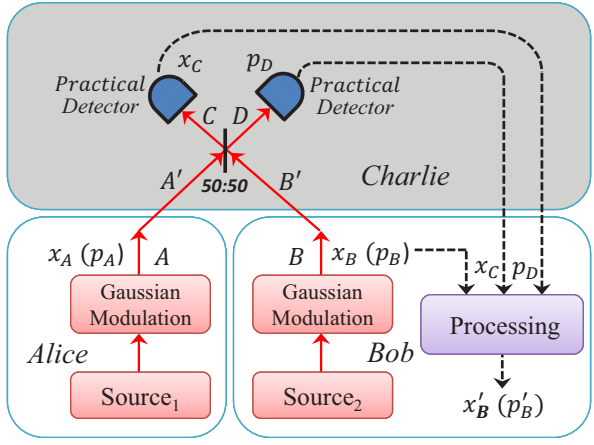}}
\caption{ (Color online) The prepare-and-measurement scheme of the CV-MDI QKD protocol using squeezed states where quantum channels and Charlie are fully controlled by Eve. Practical detectors on Charlie's side have the same quantum efficiency and electronic noise.}
\end{figure}

\emph{Step 1}. Alice and Bob randomly draw values $x_A (p_A)$ and $x_B (p_B)$ from two Gaussian distributed set with $0$ mean and variance $V_A-1$ and $V_B-1$, respectively, and use these numbers to modulate \emph{x}-quadrature (\emph{p}-quadrature) of the squeezed state. The modulation processing can be achieved by mixing the x-quadrature (p-quadrature) squeezed states on a beamsplitter of high transmissivity ($T \sim 99 \%$) with a coherent state of intensity $\frac{{x_A^2\left( {p_A^2} \right)}}{{1 - T}}$ and $\frac{{x_B^2\left( {p_B^2} \right)}}{{1 - T}}$~\cite{Patron_PhysRevLett_2009}. Then they send these states to the untrusted third party (Charlie) through two different quantum channels.

\emph{Step 2}.Charlie combines two received modes $A'$ and $B'$ with a beam splitter (50:50), and the output modes of the beam splitter are $C$ and $D$. Then he measures \emph{x}-quadrature of mode $C$ and \emph{p}-quadrature of mode $D$ by homodyne detectors and publicly announces the measurement results $x_{C}, p_{D}$ to Alice and Bob through classical channels.

\emph{Step 3}. After receiving Charlie's measurement results $x_{C}, p_{D}$, Bob modifies his data to ${x'_B}$ (${p'_B}$), while Alice keeps her data $x_A (p_A)$ unchanged.

\emph{Step 4}. Once Alice and Bob have collected a sufficiently large amount of correlated data, they first perform a parameter estimation from a randomly chosen sample of final data ${x_A}$ (${p_A}$) and ${x'_B}$ (${p'_B}$). Then Alice and Bob proceed with classical data postprocessing namely, information reconciliation and privacy amplification using an authenticated public channel. The reconciliation can be performed in two ways: either direct reconciliation (DR) or reverse reconciliation (RR).

\begin{figure}[t]
\centerline{\includegraphics[width=7cm]{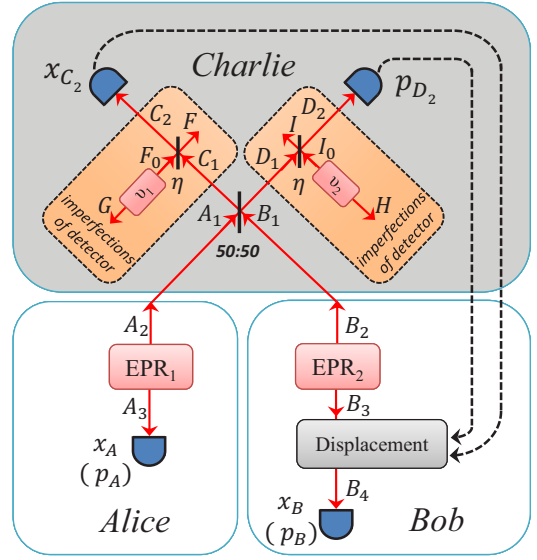}}
\caption{ (Color online) The entanglement-based scheme of the CV-MDI QKD protocol using squeezed states where all detectors represent homodyne detection and EPR states are two-mode squeezed states. Two quantum channels and Charlie are fully controlled by Eve, but Eve has no access to the apparatuses in Alice's and Bob's stations. The imperfection of the detectors is characterized by quantum efficiency $\eta$ and electronic noise $\upsilon_1 = \upsilon_2 = \upsilon_{el}$.}\label{fig2}
\end{figure}

The PM description presented above is equivalent to the EB scheme shown in Fig.~\ref{fig2}. Although the EB version does not correspond to the actual implementation, it is fully equivalent to the PM version from a secure point of view, and it provides a powerful description for establishing security proof~\cite{Grosshans_QIC_2003}. The EB scheme of the proposed CV-MDI QKD using squeezed states can be described as follows:

\emph{Step 1}. Alice and Bob respectively generate an Einstein-Podolsky-Rosen (EPR) state EPR$_1$ and EPR$_2$ with variance $V_A$ and $V_B$ and keep mode $A_3$ and $B_3$ in each side. Then they send the other mode $A_2$ and $B_2$ to the untrusted third party (Charlie) through two different quantum channels with length $L_{AC}$ and $L_{BC}$.

\emph{Step 2}.Charlie combines two received modes $A_1$ and $B_1$ with a beam splitter (50:50) and the output modes of the beam splitter are $C$ and $D$. Then he measures the \emph{x}-quadrature of mode $C$ and \emph{p}-quadrature of mode $D$ with homodyne detectors and publicly announces the measurement results $x_{C}, p_{D}$ to Alice and Bob through classical channels.

\emph{Step 3}.Bob displaces the mode $B_3$ to $B_4$ by operation $D\left(x_{C}', p_{D}'\right)$ which depends on Charlie's announced results $\left\{x_C, p_D\right\}$. Then Bob measures the mode $B_4$ to get the final data $\left\{x_B \left(p_B\right)\right\}$ using homodyne detection which randomly detects the \emph{x}-quadrature or \emph{p}-quadrature. Alice also measures the mode $A_3$ to get the final data $\left\{x_A \left(p_A\right)\right\}$ using homodyne detection.

\emph{Step 4}. Once Alice and Bob have collected a sufficiently large amount of correlated data, they use an authenticated public channel to perform parameter estimation from a randomly chosen sample of final data $\left\{x_A, p_A\right\}$ and $\left\{x_B, p_B\right\}$. Then Alice and Bob proceed with classical data postprocessing namely information reconciliation and privacy amplification to distill a secret key. The reconciliation can be performed in two ways: either direct reconciliation (DR) or reverse reconciliation (RR).

The detector's imperfection is characterized by quantum efficiency $\eta$ and electronic noise $\upsilon_{el}$, which is shown in Fig.~\ref{fig2}. The variance $\upsilon_{1,2}$ of the thermal state ${\rho _{{F_0}}}$ and ${\rho _{{I_0}}}$ is chosen to obtain the appropriate expression for practical homodyne detection $\upsilon_{1,2}  = 1 + {{{\upsilon _{el}}} \mathord{\left/ {\vphantom {{{\upsilon _{el}}} {\left( {1 - \eta } \right)}}} \right. \kern-\nulldelimiterspace} {\left( {1 - \eta } \right)}}$~\cite{Lodewyck_PhysRevA_2007}.

From the analysis above, one can find that the EB scheme proposed here shares the same demonstration with the one in Ref.~\cite{Zhengyu_PhysRevA_2013} except for that what measurements Alice and Bob use are substituted by homodyne detection. Thus, the EB scheme discussed here is equivalent to the conventional CV-QKD with squeezed states and homodyne detection. The (asymptotical) secret key rate ${K}$ against collective attacks for reverse reconciliation is given by~\cite{Devetak_ProcRSoc_2005}
\begin{align}
 {K} = \beta I\left( {A:B} \right) - \chi \left( {B:E} \right),
\end{align}
where $\beta \in[0,1]$ is the reconciliation efficiency, $I(A:B)$ is the classical mutual information between Alice and Bob, $\chi(B:E)$ is the Holevo quantity~\cite{Nielsen_QCQI}
\begin{align}
 \chi \left( {B:E} \right) = S\left( {{\rho _E}} \right) - \sum\nolimits_{x_B} {p\left( x_B \right)S\left( {{\rho _{E|x_B}}} \right)},
\end{align}
where $S(\rho)$ is the von Neumann entropy of the quantum state $\rho$, $x_B$ is Bob's measurement result obtained with the probability $p\left( x_B \right)$, $\rho _{E|x_B}$ is the corresponding state of Eve's ancillary, and ${\rho _E} = \sum\nolimits_{x_B} {p\left( x_B \right){\rho _{E|x_B}}}$ are Eve's partial states.

Firstly, Eve is able to purify the whole system $\rho_{A_3 B_4}$ to maximize her information, we have $S\left( {{\rho _E}} \right) = S\left( {{\rho _{{A_3}{B_4}}}} \right)$. Secondly, after Bob's projective measurement resulting in $x_B$, the system $\rho_{A_3 E}$ is pure, so that ${S\left( {E|x_B} \right) = S\left( {{A_3}|x_B} \right)}$. In practical experiment, we calculate the covariance matrix ${{\gamma _{{A_3}{B_4}}}}$ of correlated variables from a randomly chosen sample of measurement data. According to the Gaussian optimality theorem, we assume the final state ${{\rho _{{A_3}{B_4}}}}$ shared by Alice and Bob is Gaussian to maximize the quantum information available to Eve. Thus, the entropies $S({\rho _{{A_3}{B_4}}})$ and $\sum\nolimits_{x_B} {p\left( x_B \right)S\left( {{\rho _{{A_3}|x_B}}} \right)}$ can be calculated using the covariance matrices ${\gamma _{{A_3}{B_4}}}$ characterizing the state ${\rho _{{A_3}{B_4}}}$ and ${{\gamma _{{A_3}|x_B}}}$ characterizing the state ${{\rho _{{A_3}|x_B}}}$. So the expression for ${\chi _{BE}}$ can be further simplified as follows
\begin{align}\label{eq}
 \chi \left( {B:E} \right) = \sum\limits_{i = 1}^2 {G\left( {\frac{{{\lambda _i} - 1}}{2}} \right) - G\left( {\frac{{{\lambda _3} - 1}}{2}} \right)},
\end{align}
where $G(x) = (x + 1)\log_2 (x + 1) - x\log_2 x$, ${\lambda _{1 \left(2\right)}}$ are the symplectic eigenvalues of the covariance matrix ${{\gamma _{{A_3}{B_4}}}}$ and ${\lambda _3}$ is the symplectic eigenvalue of the covariance matrix ${\gamma _{{A_3}|x_B}}$.

As discussed above, in experiment Alice and Bob can get the covariance matrix ${\gamma _{{A_3}{B_4}}}$ from parameter estimation step. The covariance matrix ${\gamma _{{A_3}{B_4}}}$ depends on the system and the gain of the displacement, which is written as
\begin{equation}\label{eq2}
{\gamma _{{A_3}{B_4}}} = \left[ {\begin{array}{*{20}{c}}
   {{\gamma _{{A_3}}}} & {\sigma _{{A_3}{B_4}}^T}  \\
   {{\sigma _{{A_3}{B_4}}}} & {{\gamma _{{B_4}}}}  \\
\end{array}} \right] = \left[ {\begin{array}{*{20}{c}}
   {a{I_2}} & {c{\sigma _z}}  \\
   {c{\sigma _z}} & {b{I_2}}  \\
\end{array}} \right],
\end{equation}
where ${{{\rm I}_n}}$ is the $n \times n$ identity matrix and ${\sigma _z}$ = diag (1, -1). The symplectic eigenvalues ${\lambda _{1 - 2}}$ of the above matrix are given by
\begin{equation}\label{eq12}
\lambda _{1,2}^2 = \frac{1}{2}\left[ {{\rm A} \pm \sqrt {{{\rm A}^2} - 4{{\rm B}^2}} } \right],
\end{equation}
where we have used the notations
\begin{equation}
\left\{ \begin{array}{l}
 {\rm A} = {a^2} + {b^2} - 2{c^2} \\
 {\rm B} = ab - {c^2} \\
 \end{array} \right..
\end{equation}

The symplectic eigenvalues ${\lambda _3}$ of the matrix ${\gamma _{{A_3}|{x_B}}} = {\gamma _{{A_3}}} - \sigma _{{A_3}{B_4}}^T{\left( {X{\gamma _{{B_4}}}X} \right)^{-1}}{\sigma _{{A_3}{B_4}}}$ ($X$ = diag (1, 0)), after Bob's homodyne measurement, is given by
\begin{equation}
{\lambda _3^2} = a\left( {a - {{{c^2}} \mathord{\left/
 {\vphantom {{{c^2}} b}} \right.
 \kern-\nulldelimiterspace} b}} \right).
\end{equation}

\begin{figure}[t]
\centerline{\includegraphics[width=7cm]{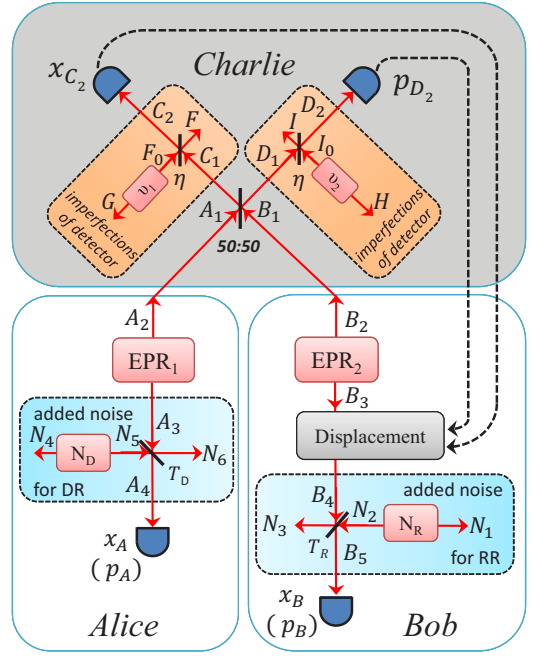}}
\caption{ (Color online) The entanglement-based scheme of the modified squeezed-state CV-MDI QKD protocol with general Gaussian added noise on the reconciliation side. Two quantum channels and Charlie are fully controlled by Eve, but Eve has no access to the apparatuses in Alice's and Bob's stations. The imperfection of the detectors is characterized by quantum efficiency $\eta$ and electronic noise $\upsilon_1 = \upsilon_2 = \upsilon_{el}$.}\label{fig3}
\end{figure}

\section{\label{sec:3} MODIFIED SQUEEZED-STATE CV-MDI QKD}
In this section, we propose a modified CV-MDI QKD protocol using squeezed states by adding proper classical Gaussian noise to the reconciliation side (Alice's side for the DR protocol or Bob's side for the RR protocol). This method has practical benefits because there exists certain preparation noise~\cite{Filip_PhysRevA_2008,Shen_JPhysB_2009,Shen_PhysRevA_2011,Weedbrook_PhysRevLett_2010,Weedbrook_PhysRevA_2012,Weedbrook_PhysRevA_2014} and detection noise~\cite{Jouguet_nature_2013,Lodewyck_PhysRevA_2007} in a practical system. If we optimize such noise in the way we discussed below, the performance of the protocol will be improved.

\begin{figure}[t]
\includegraphics[width=8.5cm]{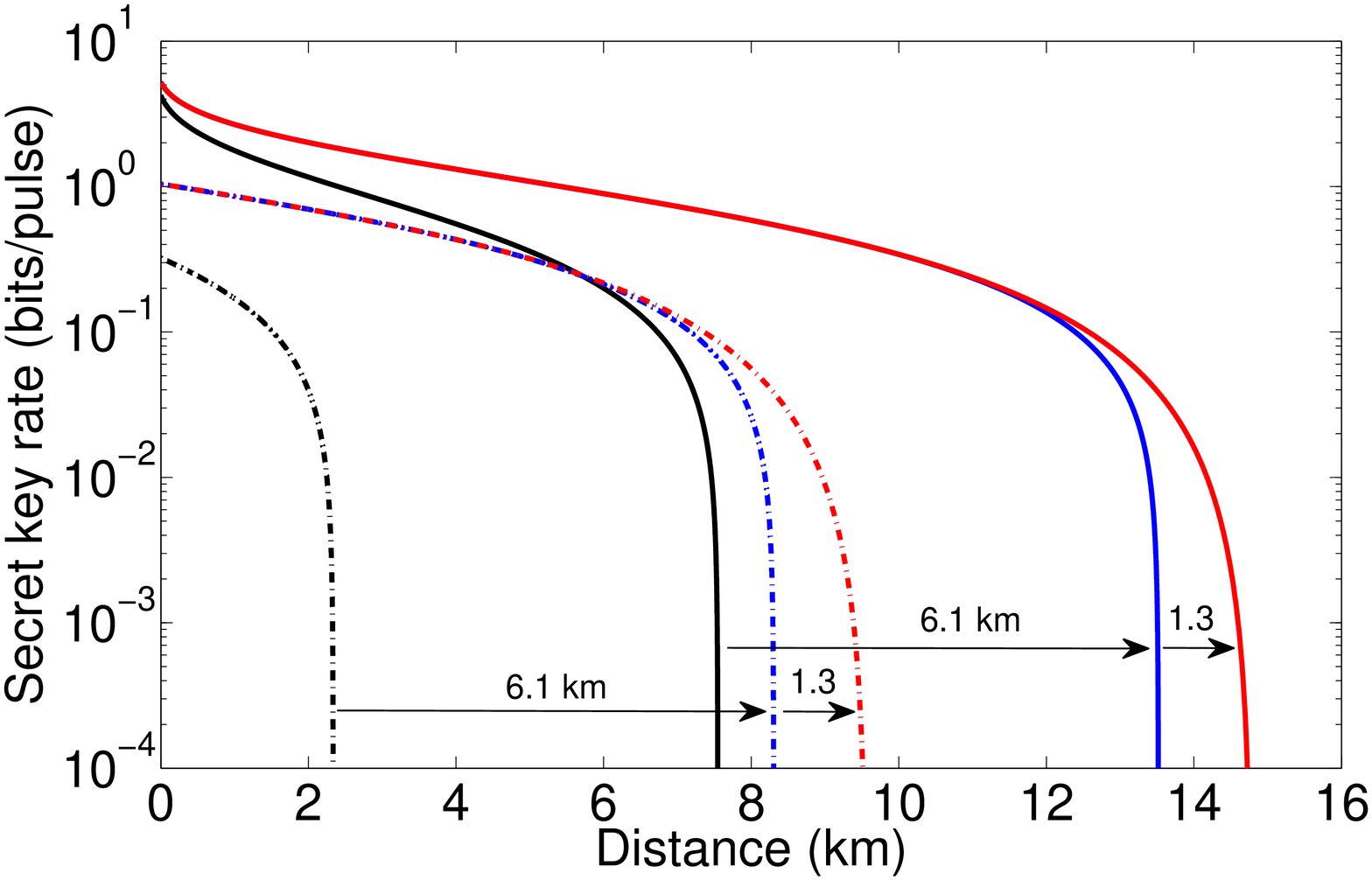}
\caption{(Color online) Secret key rates of the coherent-state (black), squeezed-state (blue) CV-MDI QKD protocol, and the modified squeezed-state CV-MDI QKD protocol (red) in the symmetric case ($L_{AC} = L_{BC}$) with perfect homodyne detectors ($\eta = 1$, ${\upsilon _{el}} = 0$) and imperfect homodyne detectors ($\eta = 0.9$, ${\upsilon _{el}} = 0.015$). The dot-dashed line and solid line represent the situation for using perfect and imperfect detectors, respectively. Here we use the ideal reconciliation efficiency $\beta = 1$, large variance $V_A = V_B = 10^5$, and $\varepsilon = 0.002$.
}\label{fig4}
\end{figure}

The EB scheme of the modified protocol is illustrated in Fig.~3, where Alice and Bob implement the original squeezed-state CV-MDI QKD protocol as we proposed in the last section but Bob adds some proper noise before his homodyne detection.

In the EB scheme, the added Gaussian phase-insensitive noise is modeled by mixing the original mode $B_4$ with a thermal state (half of an EPR) of variance $N_R$ by a beam splitter of transmissivity $T_R$, thus ${\chi _N} = {{\left( {1 - {T_R}} \right){N_R}} \mathord{\left/
 {\vphantom {{\left( {1 - {T_R}} \right){N_R}} {{T_R}}}} \right.
 \kern-\nulldelimiterspace} {{T_R}}}$. In the corresponding PM scheme, the added noise means that Bob adding proper classical Gaussian noise of variance ${\chi _N} = {{\left( {1 - {T_R}} \right){N_R}} \mathord{\left/
 {\vphantom {{\left( {1 - {T_R}} \right){N_R}} {{T_R}}}} \right.
 \kern-\nulldelimiterspace} {{T_R}}}$ before sending the modulated squeezed states to Charlie.

Then we follow the analysis of Sec.~\ref{sec:2} for collective attacks, but it is clear from Fig.~3 that two additional modes $N_1$ and $N_3$ need to be included in the calculation of the reverse reconciliation protocol. Here we only derive the expression of the reverse reconciliation protocol and for the direct reconciliation protocol we can use the similar method to calculate. By replacing the Eq.~\ref{eq}, $\chi \left( {B:E} \right)$ is calculated from the following equation:
\begin{align}
\chi \left( {B:E} \right) = \sum\limits_{i = 1}^2 {G\left( {\frac{{{\lambda _i} - 1}}{2}} \right) - \sum\limits_{i = 3}^5 {G\left( {\frac{{{\lambda _i} - 1}}{2}} \right)} } ,
\end{align}
where ${\lambda _{1 \left(2\right)}}$ remains the same as the Eq.~\ref{eq12} while ${\lambda _{3 \left(4, 5\right)}}$ represents the symplectic eigenvalues of the covariance matrix ${\gamma _{{A_3 N_1 N_3}|x_B}}$, which is given by
\begin{equation}
{\gamma _{{A_3}{N_1}{N_3}|{x_B}}} = {\gamma _{{A_3}{N_1}{N_3}}} - \sigma _{{A_3}{N_1}{N_3}{B_5}}^T{\left( {X{\gamma _{{B_5}}}X} \right)^{ - 1}}{\sigma _{{A_3}{N_1}{N_3}{B_5}}},
\end{equation}
where the matrices ${\gamma _{{A_3}{N_1}{N_3}}}$, ${\gamma _{{B_5}}}$ and $\sigma _{{A_3}{N_1}{N_3}{B_5}}$ can all be derived from the decomposition of the matrix
\begin{equation}
{\gamma _{{A_3}{N_1}{N_3}{B_5}}} = \left[ {\begin{array}{*{20}{c}}
   {{\gamma _{{A_3}{N_1}{N_3}}}} & {\sigma _{{A_3}{N_1}{N_3}{B_5}}^T}  \\
   {{\sigma _{{A_3}{N_1}{N_3}{B_5}}}} & {{\gamma _{{B_5}}}}  \\
\end{array}} \right].
\end{equation}

The above matrix can be derived with an appropriate rearrangement of lines and columns from the matrix describing the system in Fig.~3
\begin{equation}
{\gamma _{{A_3}{B_5}{N_3}{N_1}}} = {Y_{B_4 N_2}}\left( {{\gamma _{{A_3}{B_4}}} \oplus {\gamma _{{N_2}{N_1}}}} \right)Y_{B_4 N_2}^T,
\end{equation}
where ${\gamma _{{A_3}{B_4}}}$ is the same as expressed by Eq.~\ref{eq2}, ${\gamma _{{N_2}{N_1}}}$ is the standard covariance matrix of an EPR state with variance $N_R$ and ${Y_{{B_4}{N_2}}} = {I_2} \oplus {Y^{BS}} \oplus {I_2}$ where ${Y^{BS}}$ can be written by
\begin{equation}
{Y^{BS}} = \left[ {\begin{array}{*{20}{c}}
   {\sqrt {{T_R}} \cdot{I_2}} & {\sqrt {1 - {T_R}} \cdot{I_2}}  \\
   { - \sqrt {1 - {T_R}} \cdot{I_2}} & {\sqrt {{T_R}} \cdot{I_2}}  \\
\end{array}} \right].
\end{equation}

\section{\label{sec:4} NUMERICAL SIMULATION AND DISCUSSION}
In this section, the performance of the proposed and the modified squeezed-state CV-MDI QKD protocol is illustrated and compared with the coherent-state based protocol~\cite{Zhengyu_PhysRevA_2013,Pirandola_arXiv_2013}.

\begin{figure}[t]
\includegraphics[width=8.5cm]{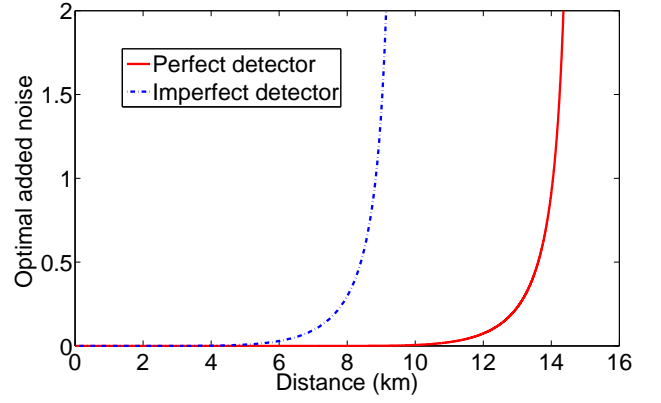}
\caption{(Color online) Optimal added noise for the modified squeezed-state CV-MDI QKD protocol in symmetric case ($L_{AC} = L_{BC}$) using perfect homodyne detectors ($\eta = 1$, ${\upsilon _{el}} = 0$) and imperfect homodyne detectors ($\eta = 0.9$, ${\upsilon _{el}} = 0.015$). Here we use the ideal reconciliation efficiency $\beta = 1$, large variance $V_A = V_B = 10^5$, and $\varepsilon = 0.002$.
}\label{fig5}
\end{figure}

As discussed above, in a practical experiment, Alice and Bob can get the covariance matrix ${\gamma _{{A_3}{B_4}}}$ from parameter estimation step. However, in a numerical simulation, a model of simulating what the channels are and what Charlie does is needed to get ${\gamma _{{A_3}{B_4}}}$. To compare with the performance of the coherent-state CV-MDI QKD protocol~\cite{Zhengyu_PhysRevA_2013}, we use the same method to simulate the channels' environment (two independent entangling cloner attacks) and Charlie's measurement (standard Bell-state measurement), which is illustrated in Fig.~\ref{fig2}. The relationships are as follows:
\begin{align}
\left\{ \begin{array}{l}
 {{\hat A}_1} = \sqrt {{T_1}} {{\hat A}_2} + \sqrt {1 - {T_1}} {{\hat E}_1} \\
 {{\hat B}_1} = \sqrt {{T_2}} {{\hat B}_2} + \sqrt {1 - {T_2}} {{\hat E}_2} \\
 {{\hat C}_1} = {{{{\hat A}_1}} \mathord{\left/
 {\vphantom {{{{\hat A}_1}} {\sqrt 2  - }}} \right.
 \kern-\nulldelimiterspace} {\sqrt 2  - }}{{{{\hat B}_1}} \mathord{\left/
 {\vphantom {{{{\hat B}_1}} {\sqrt 2 }}} \right.
 \kern-\nulldelimiterspace} {\sqrt 2 }} \\
 {{\hat D}_1} = {{{{\hat A}_1}} \mathord{\left/
 {\vphantom {{{{\hat A}_1}} {\sqrt 2  + }}} \right.
 \kern-\nulldelimiterspace} {\sqrt 2  + }}{{{{\hat B}_1}} \mathord{\left/
 {\vphantom {{{{\hat B}_1}} {\sqrt 2 }}} \right.
 \kern-\nulldelimiterspace} {\sqrt 2 }} \\
 {{\hat C}_2} = \sqrt \eta  {{\hat C}_1} + \sqrt {1 - \eta } {{\hat F}_0} \\
 {{\hat D}_2} = \sqrt \eta  {{\hat D}_1} + \sqrt {1 - \eta } {{\hat I}_0} \\
 {{\hat B}_{4x}} = {{\hat B}_{3x}} + g{{\hat C}_{2x}} \\
 {{\hat B}_{4p}} = {{\hat B}_{3p}} - g{{\hat D}_{2p}} \\
 \end{array} \right..
\end{align}

\begin{figure}[t]
\includegraphics[width=8.8cm]{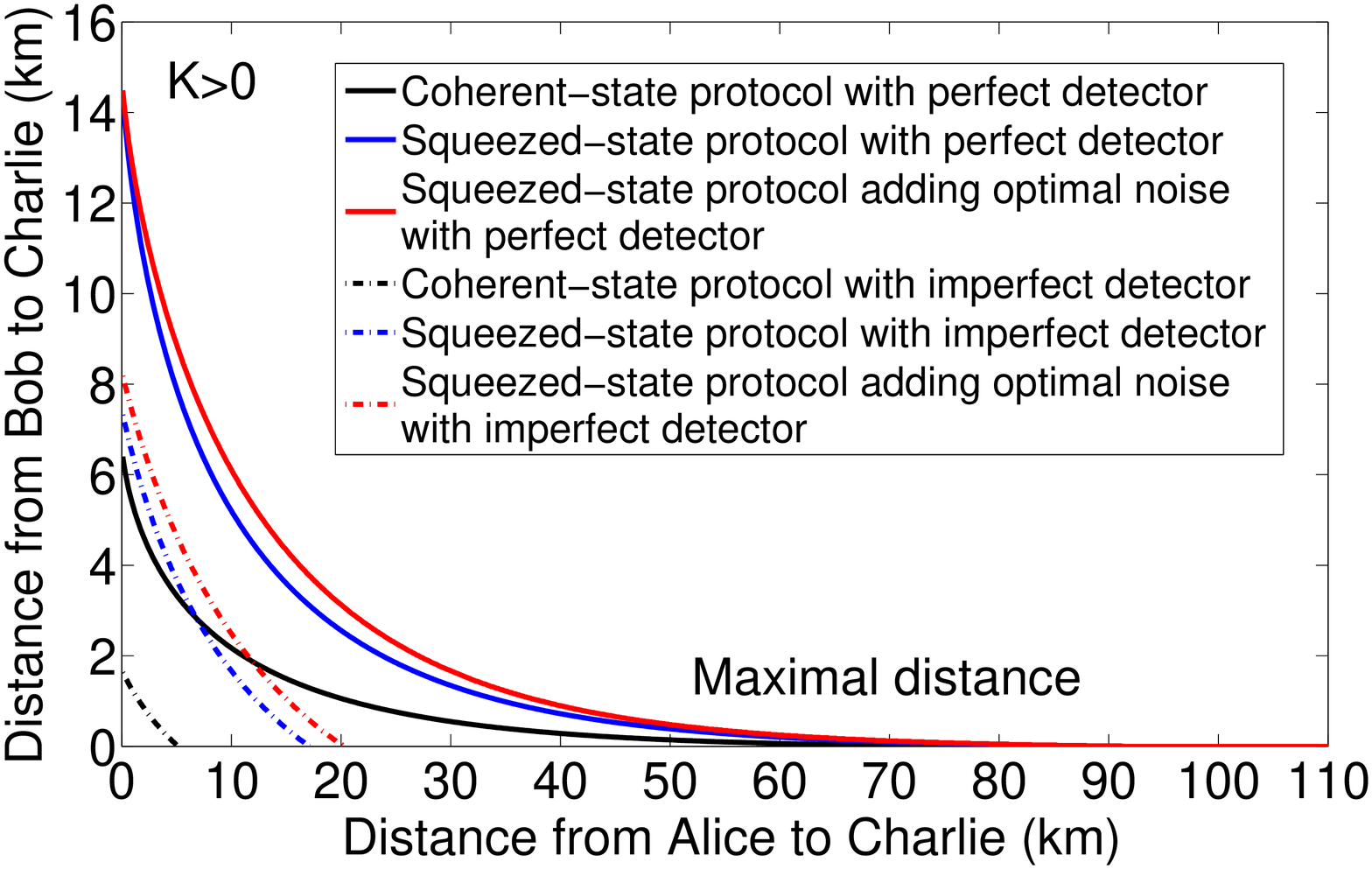}
\caption{(Color online) A comparison among the maximal transmission distance for the coherent-state, squeezed-state CV-MDI QKD protocol, and the modified squeezed-state CV-MDI QKD protocol with perfect homodyne detectors ($\eta = 1$, ${\upsilon _{el}} = 0$) and imperfect homodyne detectors ($\eta = 0.9$, ${\upsilon _{el}} = 0.015$), within which the key rate $K$ is positive. Here we use the ideal reconciliation efficiency $\beta = 1$, large variance $V_A = V_B = 10^5$, and $\varepsilon = 0.002$.
}\label{fig6}
\end{figure}

The parameters that will affect the secret key rate are the reconciliation efficiency $\beta$, the variance of Alice's and Bob's modulation $V_A-1$, $V_B-1$, the transmission efficiency $T_1$, $T_2$, excess noise $\varepsilon_1$, $\varepsilon_2$ of two quantum channels, the inefficiency $\eta$ and the electronic noise ${\upsilon _{el}}$ of the practical homodyne detector. Here we first choose large variance $V_A = V_B = 10^5$ to see the performance of ideal modulation, then we will use the practical variance $V_A = V_B = 5.04$ to see the realistic performance. Excess noises are $\varepsilon_1 = \varepsilon_2 = \varepsilon = 0.002$ and transmittances are $T_1 = 10^{-\alpha L_{AC}/10}$, $T_2 = 10^{-\alpha L_{BC}/10}$ ($\alpha = 0.2$~dB/km) for simulation, which are standard parameters in one-way CV-QKD experiment~\cite{Jouguet_nature_2013}. Furthermore, we use $\eta = 1$, ${\upsilon _{el}} = 0$ representing for the perfect homodyne detector and $\eta = 0.9$, ${\upsilon _{el}} = 0.015$ for the imperfect detector.

\begin{figure}[t]
\includegraphics[width=8.8cm]{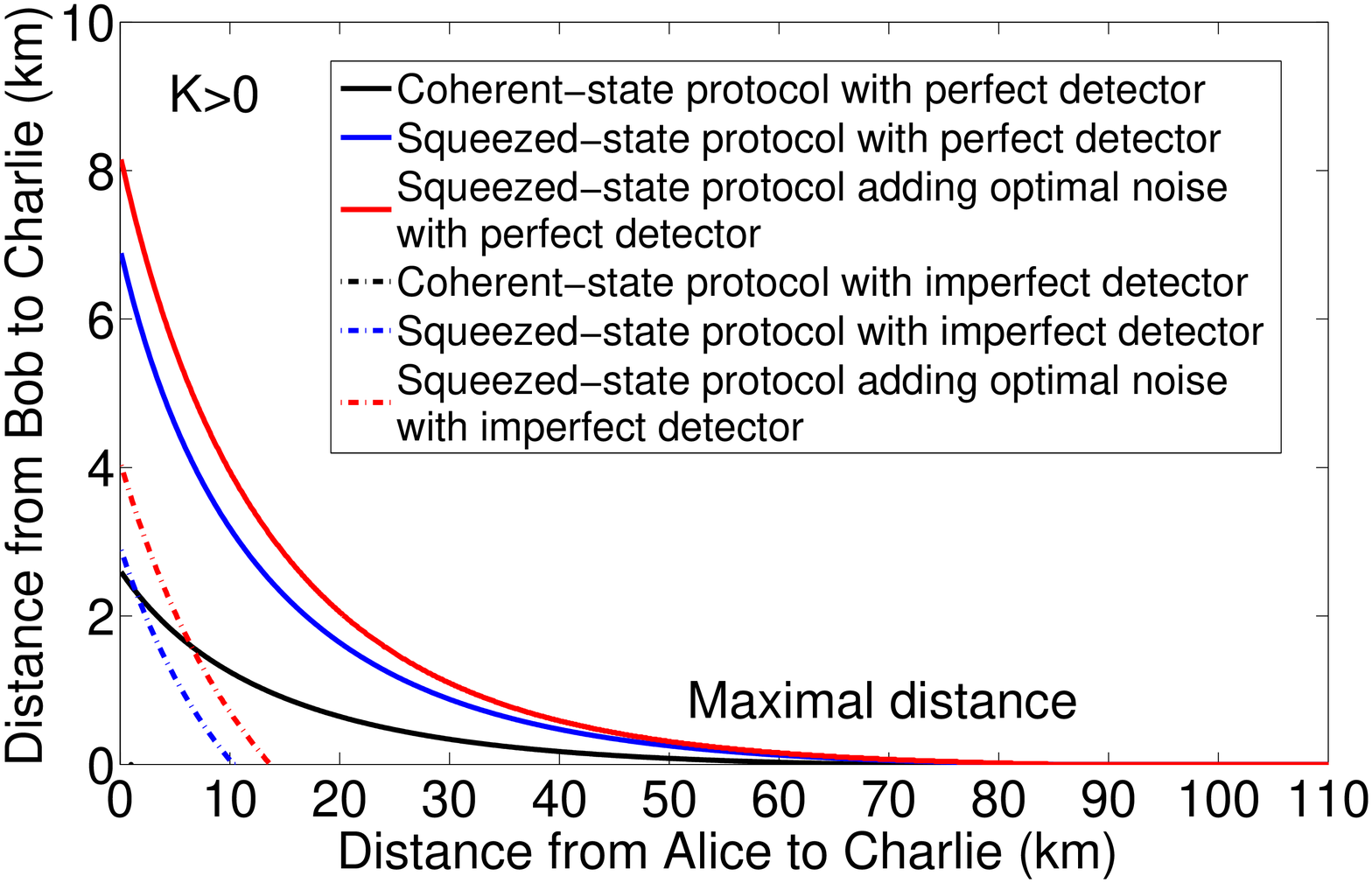}
\caption{(Color online) A comparison among the maximal transmission distance for the coherent-state, squeezed-state CV-MDI QKD protocol, and the modified squeezed-state CV-MDI QKD protocol with perfect homodyne detectors ($\eta = 1$, ${\upsilon _{el}} = 0$) and imperfect homodyne detectors ($\eta = 0.9$, ${\upsilon _{el}} = 0.015$), within which the key rate $K$ is positive. Here we use the realistic variance $V_A = V_B = 5.04$ and $\varepsilon = 0.002$.
}\label{fig7}
\end{figure}

\begin{figure}[b]
\includegraphics[width=8.5cm]{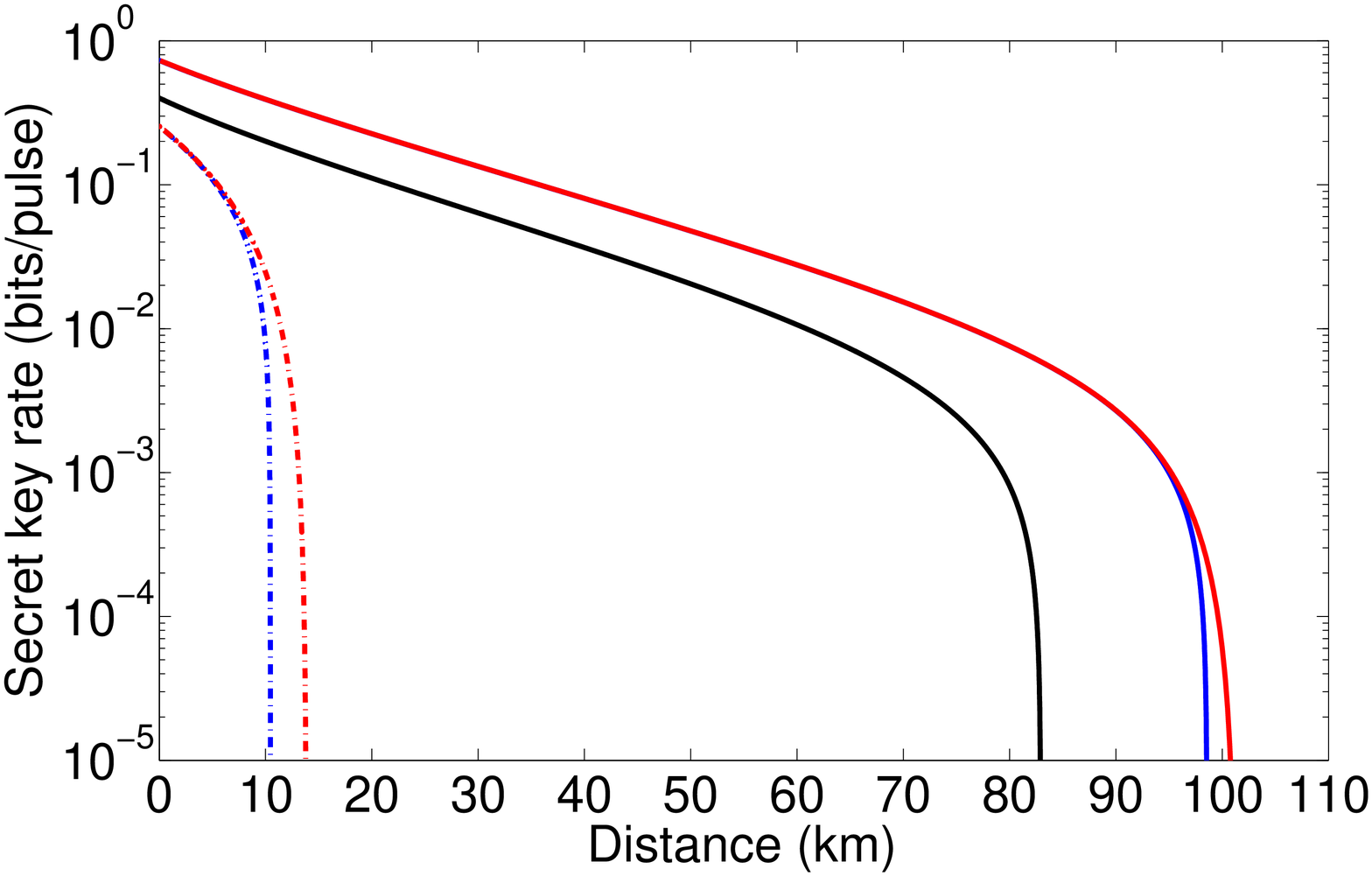}
\caption{(Color online) Secret key rates of the coherent-state (black), squeezed-state (blue) CV-MDI QKD protocol, and the modified squeezed-state CV-MDI QKD protocol (red) in the most asymmetric case ($L_{BC} = 0 km$) with perfect homodyne detectors ($\eta = 1$, ${\upsilon _{el}} = 0$) and imperfect homodyne detectors ($\eta = 0.9$, ${\upsilon _{el}} = 0.015$). The dot-dash line and solid line represent the situation of using perfect and imperfect detectors, respectively. Here we use the realistic parameters: $V_A = V_B = 5.04$ and $\varepsilon = 0.002$.
}\label{fig8}
\end{figure}

First, we consider the performance of the symmetric case where the length of two quantum channels is equal ($L_{AC} = L_{BC}$). We calculate the secret key rate $K$ as a function of transmission distance $d = L_{AC} = L_{BC}$ with perfect detectors or imperfect detectors. The simulation results are shown in Fig.~\ref{fig4}, where we also calculate the CV-MDI-QKD protocol using coherent states~\cite{Zhengyu_PhysRevA_2013} for comparison. We find that the secret key rate of the squeezed-state CV-MDI QKD protocol is always larger than coherent-state based protocol. Explicitly, the total maximal transmission distance ($L_{AB} = L_{AC}+L_{BC}$) of our squeezed-state CV-MDI QKD protocol increases both $6.1$~km using perfect detectors and imperfect detectors than that of the coherent-state based protocol. Furthermore, by adding proper Gaussian noise in Bob's side, the performance of the modified squeezed-state CV-MDI QKD protocol improves. The total maximal transmission distance increases both $1.3$~km and the optimal added noise is illustrated in Fig.~\ref{fig5}. We also observe that the imperfections of the homodyne detectors decrease $L_{AB}$, i.~e.~, using $\eta = 0.9$, ${\upsilon _{el}} = 0.015$ homodyne detectors decrease $L_{AB}$ $7.2$~km for the modified squeezed-state CV-MDI QKD protocol. This is because the imperfections of the homodyne detectors dramatically increase the equivalent excess noise of the whole system~\cite{Zhengyu_PhysRevA_2013}.

Second, we also find that when Charlie's position is close to Bob, the total maximal transmission distance $L_{AB}$ will increase to a relatively longer distance. To understand this phenomenon intuitively, we can treat the EB scheme of the CV-MDI QKD protocol as a CV teleportation process, in which Bob and Charlie preshare an EPR source and then teleport mode $A¡¯$ from Alice to Bob. Any loss and noise in Bob¡¯s channel will decrease the quality of the EPR source and will decrease the fidelity of the teleportation result. Thus, here we also consider the performance of the asymmetric case where $L_{AC} \neq L_{BC}$. As illustrated in Fig.~\ref{fig6}, $L_{AB}$ increases when $L_{BC}$ decreases. In the asymmetric case, the performance of the modified squeezed-state CV-MDI QKD protocol is also improved by adding the proper noise to Bob's side. The optimal input noise can enhance $L_{AB}$ to $91.2$~km when using perfect detectors and $20.0$~km when using imperfect detectors. Then we change the variance from ideal ($V_A = V_B = 10^5$) to a practical one ($V_A = V_B = 5.04$), which is shown in Fig.~\ref{fig7}. The total maximal transmission distance $L_{AB}$ can also reach $84.4$~km using perfect detectors and $13.7$~km using imperfect detectors, which allows one to directly use the EPR state as the source in a practical experiment. And if Alice and Bob use EPR sources, the squeezed-state CV-MDI QKD protocol will be more secure because they could completely outplay side-channel attacks in their private spaces~\cite{Pirandola_PhysRevLett_2012_MDI}. Compared with the performance of using perfect and imperfect detectors, we find that the inefficiency and electronic noise of the practical homodyne detector have a significant influence on the transmission distance. It is possible that these imperfections can be compensated by optical phase-sensitive-amplifiers~\cite{Fossier_JPhysB_2009,My_JPhysB_2014}.

\begin{figure}[t]
\includegraphics[width=8.5cm]{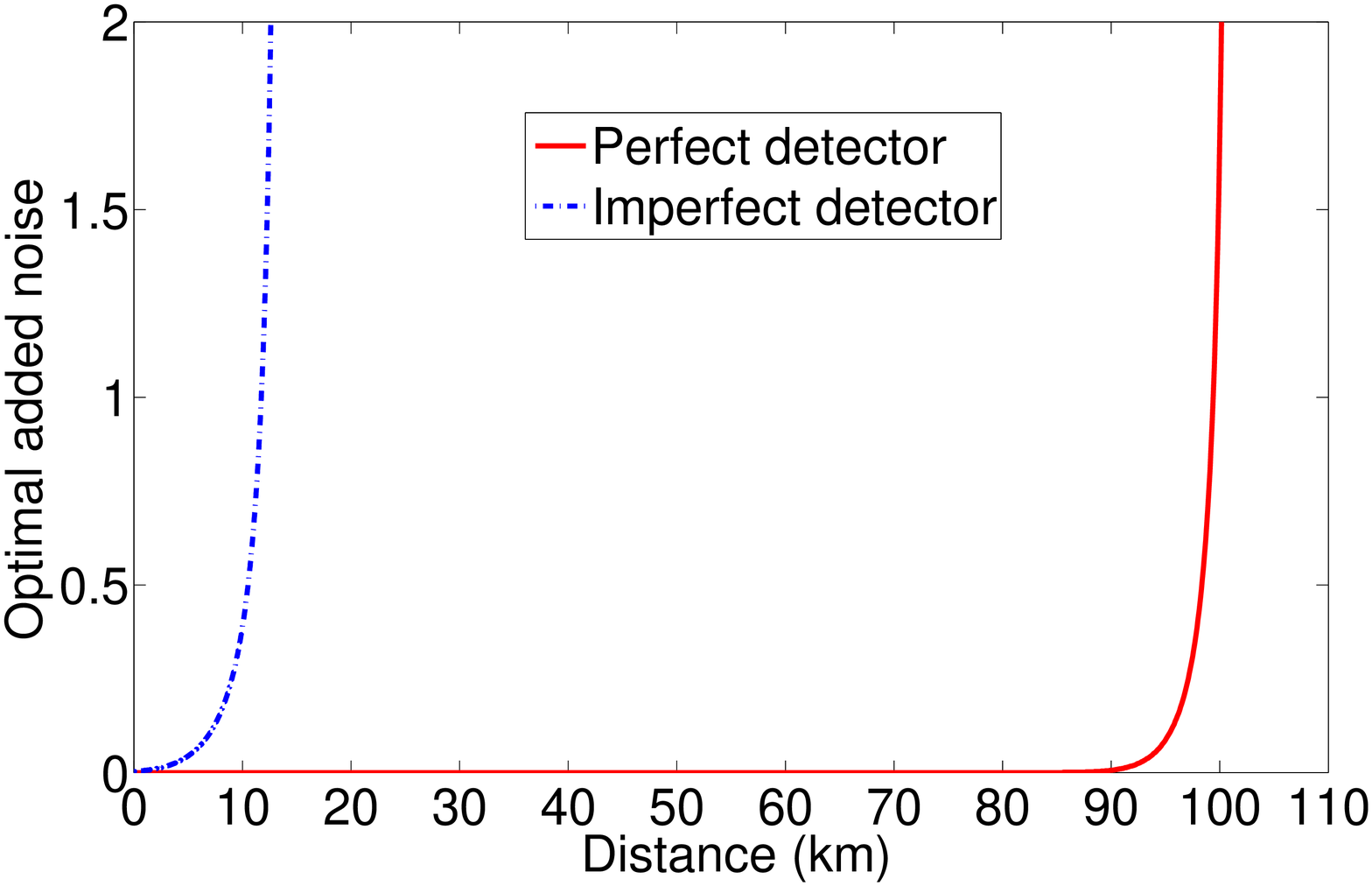}
\caption{(Color online) Optimal input noise for the modified squeezed-state CV-MDI QKD protocol in the most asymmetric case ($L_{BC} = 0 km$) using perfect homodyne detectors ($\eta = 1$, ${\upsilon _{el}} = 0$) and imperfect homodyne detectors ($\eta = 0.9$, ${\upsilon _{el}} = 0.015$). Here we use the realistic parameters: $V_A = V_B = 5.04$ and $\varepsilon = 0.002$.
}\label{fig9}
\end{figure}

Finally, we consider the performance of the most asymmetric case where we put Charlie on Bob's side ($L_{BC} = 0 km$). As illustrated in Fig.~\ref{fig8}, $L_{AB}$ increases dramatically compared with the symmetric case. In the most asymmetric case, the secret key rate of the squeezed-state CV-MDI QKD protocol is always larger than that of the coherent-state based protocol. The transmission distance of the coherent-state protocol is zero. Thus, there is no line representing the coherent-state CV-MDI QKD protocol using imperfect detectors in Fig.~\ref{fig8}. Thus, the total maximal transmission distance increases $15.7$~km and $10.5$~km when using perfect detectors and imperfect detectors. Furthermore, the modified squeezed-state CV-MDI QKD protocol achieves optimal performance than the original protocol by adding proper noise in Bob's side, i.~e.~, $L_{AB}$ increases more $4.9$~km and $3.3$~km. The optimal input noise is illustrated in Fig.~~\ref{fig9}. The reason of the improvement is that the added noise not only lowers the mutual information between Alice and Bob, but also lowers that between Eve and Bob. When the influence on Eve and Bob is stronger than that on Alice and Bob, the secret key rate is enhanced and the performance is improved.

\section{\label{sec:5}CONCLUSION}
In this paper, we proposed a continuous-variable measurement-device-independent QKD protocol using squeezed states, which outperforms the coherent-state-based protocol in terms of the secret key rate and maximal transmission distance. Security analysis shows that the protocol is immune to attacks against the detector and secure against collective attacks in the asymptotical limit. Furthermore, we also presented a method to optimize the performance of the reverse reconciliation squeezed-state continuous-variable measurement-device-independent QKD protocol by adding proper Gaussian noise to Bob's side. It is found that there is an optimal noise Bob need to add to maximize the secret key rate and total transmission distance for the reverse reconciliation protocol. The resulting protocol exhibits the optimal performance and shows the potential of long-distance secure communication using the continuous-variable measurement-device-independent QKD protocol. We restricted our discussion to collective-attack cases in this paper. Since the protocol using squeezed states and homodyne detection is also shown to be secure against coherent attacks in the finite-size regime~\cite{Furrer_PhysRevLett_2012,Furrer_arXiv_2014}, an interesting extension to this paper would be to further derive the security bound against coherent attacks.

\begin{acknowledgments}
This work was supported in part by the National Basic Research Program of China (973 Pro-gram) under Grant 2012CB315605 and 2014CB340102, in part by the National Science Fund for Distinguished Young Scholars of China (Grant No. 61225003), in part by the National Natural Science Foundation under Grant 61101081, 61271191 and 61271193, in part by the Fund of State Key Laboratory of Information Photonics and Optical Communications, and in part by the Fundamental Research Funds for the Central Universities.
\end{acknowledgments}


\begin{thebibliography}{99}
\bibitem{Text}
In experimental papers, squeezing is often measured in decibels, defined so that a squeezing degree $r$ corresponds to $10{\log _{10}}\left( {{e^{2r}}} \right)dB$. Thus, 10 dB corresponds to $r = 1.15$, i.e., to an EPR variance $V = \cosh \left( {2r} \right) = 5.04$.
\end{thebibliography}
\end{document}